\documentclass[aps,prd,lengthcheck,twocolumn,showpacs,superscriptaddress,groupedaddress]{revtex4-1}

\usepackage{graphicx}
\usepackage{dcolumn}
\usepackage{bm}        
\usepackage{amssymb}   

\begin{document}
\newcommand{\MNRAS}{Mon. Not. R. Astron. Soc.}
\newcommand{\apjl}{Astrophys. J. Lett.}
\newcommand{\grl}{Geophys. Res. Lett.}
\newcommand{\jgr}{J. Geophys. Res.}
\newcommand{\apjs}{Astrophys. J. Supp.}
\newcommand{\aap}{Astron. Astrophys.}

\title{Fast Diffusion of Magnetic Field in Turbulence and Origin of Cosmic Magnetism} 
\author{Jungyeon Cho}
\email[]{jcho@cnu.ac.kr}
\altaffiliation{Dept. of Astronomy \& Space Science, Chungnam National Univ., Daejeon, Korea}
\date{\today}

\begin{abstract}
Turbulence is believed to play important roles in
the origin of cosmic magnetism.
While it is well known that turbulence can efficiently
amplify a uniform or spatially homogeneous seed magnetic field,
it is not clear whether or not we can draw a similar conclusion
for a localized seed magnetic field.
The main uncertainty is the rate of magnetic field diffusion on
scales larger than the outer scale of turbulence. 
To measure the diffusion rate of magnetic field on those large scales,
we perform a numerical simulation in which
the outer scale of turbulence is much smaller than
the size of the system.
We numerically compare diffusion of a localized seed magnetic field 
and a localized passive scalar.
We find that diffusion of the magnetic field can be much faster than 
that of the passive scalar and that turbulence can efficiently
amplify the localized seed magnetic field. 
Based on the simulation result, we construct a model for fast diffusion of magnetic field.
Our model suggests that a localized seed magnetic field  can fill the whole system in $\sim L_{sys}/L$ times 
the large-eddy turnover time and that
growth of the magnetic field stops in $\sim \max( 15, L_{sys}/L )$
times the large-eddy turnover time,
where $L_{sys}$ is the size of the system and $L$ is the driving scale.
Our finding implies that, regardless of the shape of the seed field,
fast magnetization is possible in turbulent systems, such as large-scale
structure of the universe or galaxies.

\end{abstract}
\maketitle

\section{Introduction}

Magnetic fields 
are ubiquitous in the universe
(see, e.g., \cite{Kro94, KulZ08}). 
However, origin of magnetic fields in the 
universe is still an unsolved problem.
The origin of cosmic magnetism can
be split into two parts - the origin of seed fields and their
amplification.
In this paper, we are mainly concerned with the latter in the presence of turbulence.

Two extreme types of seed fields can exist - spatially homogeneous
and spatially localized ones.
If the seed magnetic fields have cosmological origins, it is likely that
their coherence lengths are larger than the size of galaxy clusters \cite{Rees87,BraEO96}
 and we can treat them as spatially uniform.
It is well known that turbulence can efficiently amplify
such a uniform seed field \cite{Bat50,  Kaz68, KulA92, Kul97,ChoV00, Sch04, BraS05, SubSH06, SchC07, Ryu08, ChoVB09}. 
When we introduce a weak \textit{uniform} (or \textit{homogeneous}) magnetic field in a turbulent medium, 
amplification of the field happens in three stages 
(see \cite{SchC07}; see also \cite{ChoV00, ChoVB09}). 
(1) Exponential growth: Stretching of
magnetic field lines occurs most actively near the velocity dissipation
scale (i.e.~the Kolmogorov scale) first, and the magnetic energy grows exponentially. 
Note that eddy turnover time is shortest at the scale.
(2) Linear growth: The exponential growth stage ends when the magnetic energy
becomes comparable to the kinetic energy at the dissipation
scale. The subsequent stage is characterized by a linear growth
of magnetic energy and a gradual increase of the stretching scale.
(3) Saturation: The amplification of magnetic
field stops when the magnetic energy density becomes comparable
to the kinetic energy density and a final, statistically steady, saturation stage
begins.

On the other hand, if the seed fields are ejected from astrophysical bodies,
such as active galactic nuclei or galaxies,
they will be highly localized in space.
Recently Cho \& Yoo \cite{ChoY12} 
showed that turbulence can also efficiently disperse and
amplify a localized seed magnetic field in an extreme case that
the outer scale of turbulence is
comparable to the size of the system.
However, 
since their results are valid only for the case the outer scale of turbulence is
comparable to the size of the system,
their result does not guarantee fast magnetic diffusion in general circumstances.

In this paper, we investigate whether diffusion of magnetic field in turbulence is in general fast.
For this purpose, we first perform a numerical simulation in which the outer scale of turbulence is very small.
If the outer scale of turbulence is much smaller than the size of the system,
it is not clear whether  
diffusion of magnetic field is fast on scales larger than
the outer scale of turbulence.
If we consider a passive scalar field, we know that diffusion of the scalar field 
on scales larger than the outer scale of turbulence
is slow  because its diffusion
over uncorrelated eddies is slow.  
In this paper, we compare diffusion of a magnetic field and a passive scalar field on scales
larger than the outer scale and
show that diffusion of the former can be much faster.
Based on the simulation, we construct a physical model for fast diffusion of magnetic field in turbulence
and apply the model for the large-scale structure of the universe.

We describe numerical method in Section II and theoretical consideration in Section III.
We present results in Section IV and discussion and summary in Section V.

\begin{figure} 
\includegraphics[width=.45\textwidth]{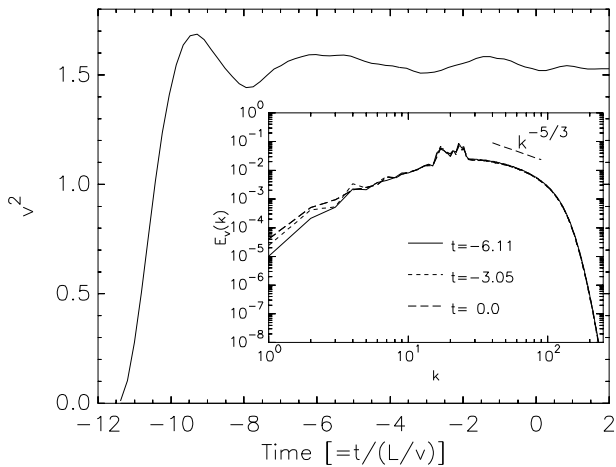}  
\caption{Evolution of turbulence before the seed magnetic field is switched on.
    We start a hydrodynamic turbulence simulation at t$\sim$-12.
    We run the simulation without a magnetic field for more than
    10 large-eddy turnover times. Then we inject a localized seed magnetic field at t=0.
    The inset shows kinetic energy spectra at 3 time points.}
\label{fig:hydro}
\end{figure}

\begin{figure*} 
\includegraphics[width=.60\textwidth]{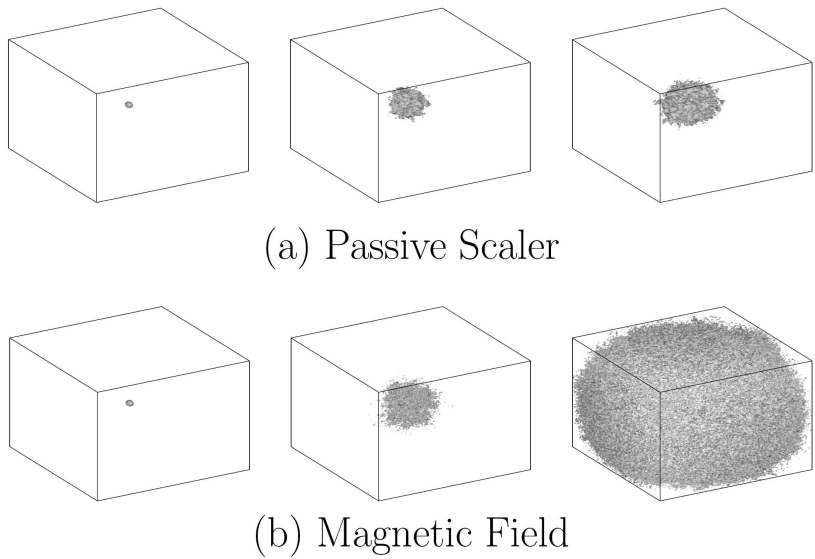}  
\caption{Diffusion of the passive scalar field (upper panels) and the  
    magnetic field (lower panels)
    at t=0, 11.5, and 28.5 (from left to right). At t=0, the fluid is fully turbulent and 
    the magnetic field and the passive scalar field exist
    only at the central part of the simulation box. 
    The driving scale ($L$) is about 1/20 of a side of the box ($L_{sys}$).}
\label{fig:diffusion}
\end{figure*}

\section{Numerical Method}

\textit{Numerical code} ---
We use a pseudospectral code to solve the 
incompressible magnetohydrodynamic (MHD) equations in a periodic box of size $2\pi$ ($\equiv L_{sys}$):
\begin{eqnarray}
\frac{\partial {\bf v} }{\partial t} =(\nabla \times {\bf B})
        \times {\bf B} -(\nabla \times {\bf v}) \times {\bf v}
      + \nu \nabla^{2} {\bf v} + {\bf f} + \nabla P^\prime ,
        \label{veq}  \hspace{5mm} \\ 
\frac{\partial {\bf B}}{\partial t}= 
     \nabla \times ({\bf v} \times{\bf B}) + \eta \nabla^{2} {\bf B} ,
     \label{beq}  \hspace{10mm}
\end{eqnarray}
where 
$ 
      \nabla \cdot {\bf v} =\nabla \cdot {\bf B}= 0,
$ 
$\bf{f}$ is random driving force,
$P'\equiv P + v^2/2$, ${\bf v}$ is the velocity,
and ${\bf B}$ is magnetic field divided by $(4\pi \rho)^{1/2}$.
We also solve the continuity equation for a passive scalar field.
We use 100 forcing components in the wavenumber range $15 < k <26$, which means the driving scale $L$ is
$\sim L_{sys}/20$.
In our simulation, 
$v\sim$1.2 and $\sim$1 before and after saturation, 
respectively (see Figure \ref{fig:3plots}(d)). 
Therefore, 
the large-eddy turnover time,  
$\sim L/v$, is approximately $\sim$0.26 and $\sim$0.31, respectively.  
In what follows, we represent time in units of $L/v$ before saturation.
Other variables have their usual meaning.
We use $512^3$ collocation points.

We use hyperviscosity and hyperdiffusion for dissipation terms. 
The power of hyperviscosity
is set to 3, such that the dissipation term in Equation (\ref{veq}) 
is replaced with
$ 
 -\nu_3 (\nabla^2)^3 {\bf v}.
$
The same expression is used for the magnetic dissipation term with $\eta_3=\nu_3$
Therefore, the magnetic Prandtl number ($=\nu/\eta$) is one.
Since we use hyperviscosity and hyperdiffusion, 
dissipation of both fields is negligible for small wavenumbers and 
it abruptly increases near $(2/3)k_{max}$, where $k_{max}$=256 \footnote{
 The abrupt increase of dissipation near k$\sim$170 causes a steep
 decrease of the energy
 spectrum for k$\gtrsim$170  (see the inset of Figure \ref{fig:hydro}).
 Note that, if there are little powers in Fourier modes with
 k $> (2/3)k_{max}$ ($\sim$170), the aliasing errors which are
 present in pseudospectral codes
 can be small.
}.

\textit{Initial conditions} ---
\textbf{
We inject a localized seed magnetic field in a turbulent medium at t=0.
In terms of the numerical process, this is achieved in the following way.
First, we start off a hydrodynamic turbulence simulation at t$\sim$-12 
with zero velocity field (Figure \ref{fig:hydro}).
Then, we run the simulation
 without the magnetic field until it reaches a steady state.
The kinetic energy density and spectra in Figure \ref{fig:hydro} 
show that the system has reached a statistically steady state before t=0.
At t=0, the localized seed magnetic field gets ``switched on''.
}

At $t=0$, 
the magnetic field has a doughnut shape, which mimics a magnetic field ejected from
an astrophysical body. 
We use the following expression for the magnetic field at t=0:
\begin{equation}
  {\bf B}(\Delta x, r_{\bot}) 
  = \frac{ B_{max} }{ 2 \sigma_0^2 e^{-1}}  r_{\bot} ^2 e^{ -r_{\bot}^2 /2 \sigma_0^2 } 
      e^{ -\Delta x^2/8 \sigma_0^2 } 
     \hat{\bf \theta}_{\perp},   \label{eq:b_shape}
\end{equation}
where $B_{max}=0.01$, $\sigma_0=4\sqrt{2}$, $r_{\bot}=( \Delta y^2+ \Delta z ^2 )^{1/2}$, 
and $\Delta x, \Delta y,$ and $\Delta z$ are distances measured from the center of the
numerical box in grid units. The unit vector $ \hat{\theta}_{\perp}$ is perpendicular to
$(\Delta x,0,0)$ and $(0,\Delta y, \Delta z)$.
Note that the maximum strength of the magnetic field at t=0 is $B_{max}$.
Since $\sigma_0=4\sqrt{2}$ in Equation (\ref{eq:b_shape}), the size of
the magnetized region at t=0 is $\sim$16 in grid units, which is
$\sim$1/32 of the simulation box size. 
Therefore, in a cluster of size $\sim$1Mpc, the size of the initially magnetized region corresponds 
to  $\sim$30kpc.

\section{Theoretical considerations}

At t=0, the magnetized region ($\sim$16 in grid units) is 
smaller 
than an outer-scale eddy ($\sim$25 in grid units).  
Since the initial magnetic field is very weak, the magnetic field will be passively
advected by turbulent motions inside an outer-scale eddy.
Therefore, on scales smaller than the outer scale of turbulence, 
evolution of the magnetic field is expected to be
very similar to that of a passive scalar.
Indeed, the standard deviation $\sigma$ of the magnetic field
distribution follows the Richardson's law ($\sigma \propto t^{3/2}$)
on scales smaller than the outer scale \cite{ChoY12}.

On scales larger than the outer scale of turbulence, the passive scalar
will diffuse over uncorrelated eddies. Therefore, we expect
that $\sigma \propto t^{1/2}$.
Then, will the magnetic field follow the same law?
This is the question we try to answer in this paper.

\section{Results}

\textit{Results} ---
Figure \ref{fig:diffusion} shows distribution of the passive scalar (upper panels) 
and $B^2$ (lower panels) at t=0, 11.5, and 28.5.
At t=0, we set the value of the passive scalar to $B^2({\bf x})$.
Therefore, both the passive scalar and $B^2$ have the same shape at t=0.
In the figure we show the regions that satisfy
$ 
    s({\bf x}) \geq 0.6 \bar{s}_c,
$ 
where $\bar{s}_c $ is the average value of $s({\bf x})$ 
inside a sphere of radius 12 (in grid units)
at the center of the simulation box.
Here $s({\bf x})$ stands for either the passive scalar or $B^2$ at ${\bf x}$.

\begin{figure}
\center
\includegraphics[angle=0,width=0.95\columnwidth]{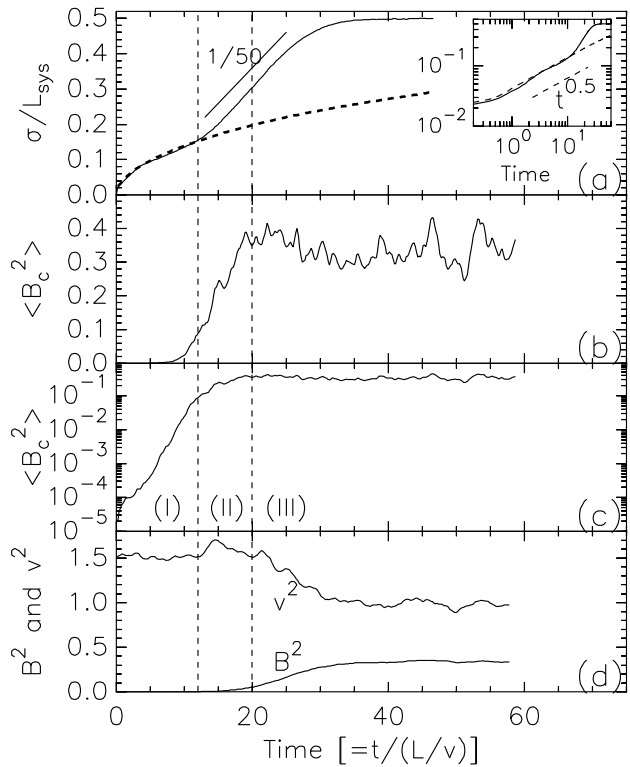}  
\caption{(a) The standard deviation ($\sigma$) of the passive scalar (dashed line) and
     the magnetic field (solid line). 
     Note that $\sigma$ of the magnetic field shows linear increase for $t\gtrsim 12$.
     (b) The average of $B^2$ at the central part of the simulation box.
     The vertical dashed lines roughly delineate three stages of 
     the magnetic field growth at the center:
     It follows in turn exponential (I), 
      linear (II) and saturation (III) stages, 
     just as a uniform magnetic
     field does. Note that it reaches the saturation stage in 15-20 large-eddy turnover times.
     (c) Same as (b). The vertical axis in drawn in a logarithmic scale.
     (d) Time evolution of $v^2$ and $B^2$.
 }
\label{fig:3plots}
\end{figure}

At t=11.5, the magnetic field (lower-middle panel) seems to have a wider distribution
than the passive scalar (upper-middle panel).
However, calculation shows that
their standard deviations are virtually same (see Figure \ref{fig:3plots}(a); compare two curves at t=11.5).
The reason the magnetic field looks  wider 
than the passive scalar is that the magnetic field has spatially  more intermittent structures.
That is, the shaded regions for the magnetic field are more porous than those for the passive scalar.
At t=28.5, the magnetic field (lower-right panel) clearly has a much wider distribution
than the passive scalar (upper-right panel).
Figure \ref{fig:3plots}(a) shows that the standard deviation of the magnetic field is indeed
larger than that of the passive scalar at t=28.5.

Figure \ref{fig:3plots}(a) shows that the passive scalar (dashed line) and the magnetic field (solid line) 
spread similarly until $t\sim$12. 
The inset shows that both of them follow the $t^{1/2}$ law for $t \lesssim 12$.
Note that, since 
the outer scale is not much larger than the initially magnetized region ($\sim$16 in grid units),
we are not able to observe the Richardson's law: we only observe the $t^{1/2}$ law that happens
when the fields diffuse over uncorrelated eddies.
Then, at $t\sim$12 the behavior of the two fields suddenly diverges.
The passive scalar continues to follow the $t^{1/2}$ law.
But, the magnetic field follows a $t^{1}$ law after $t\sim$12.

The slope of $\sigma(t)/L_{sys}$ of magnetic field after $t>12$ in Figure \ref{fig:3plots}(a) is $\sim$1/50.
Therefore we can write $\sigma(t)/L_{sys} \sim (1/50)[t/(L/v)]$, or
\begin{equation}
  \sigma(t) / L  
                          \sim 0.4 t/( L/v ),
\end{equation}
which implies 
that the spreading speed of 
the magnetized region (size $\sim$2$\sigma$) is approximately $v$.

\begin{figure*}
\includegraphics[angle=0,width=0.30\textwidth]{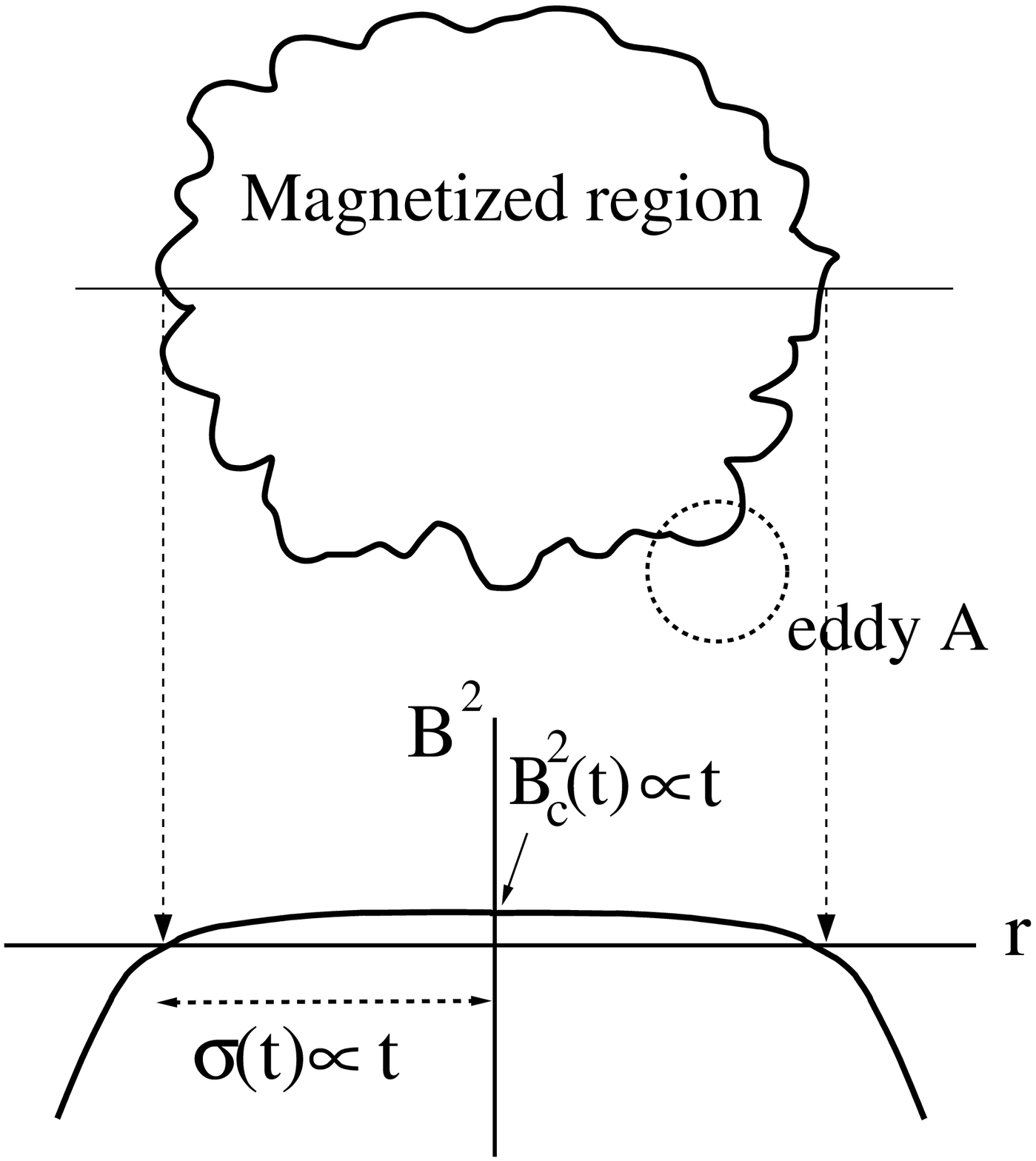}   
\includegraphics[angle=0,width=0.34\textwidth]{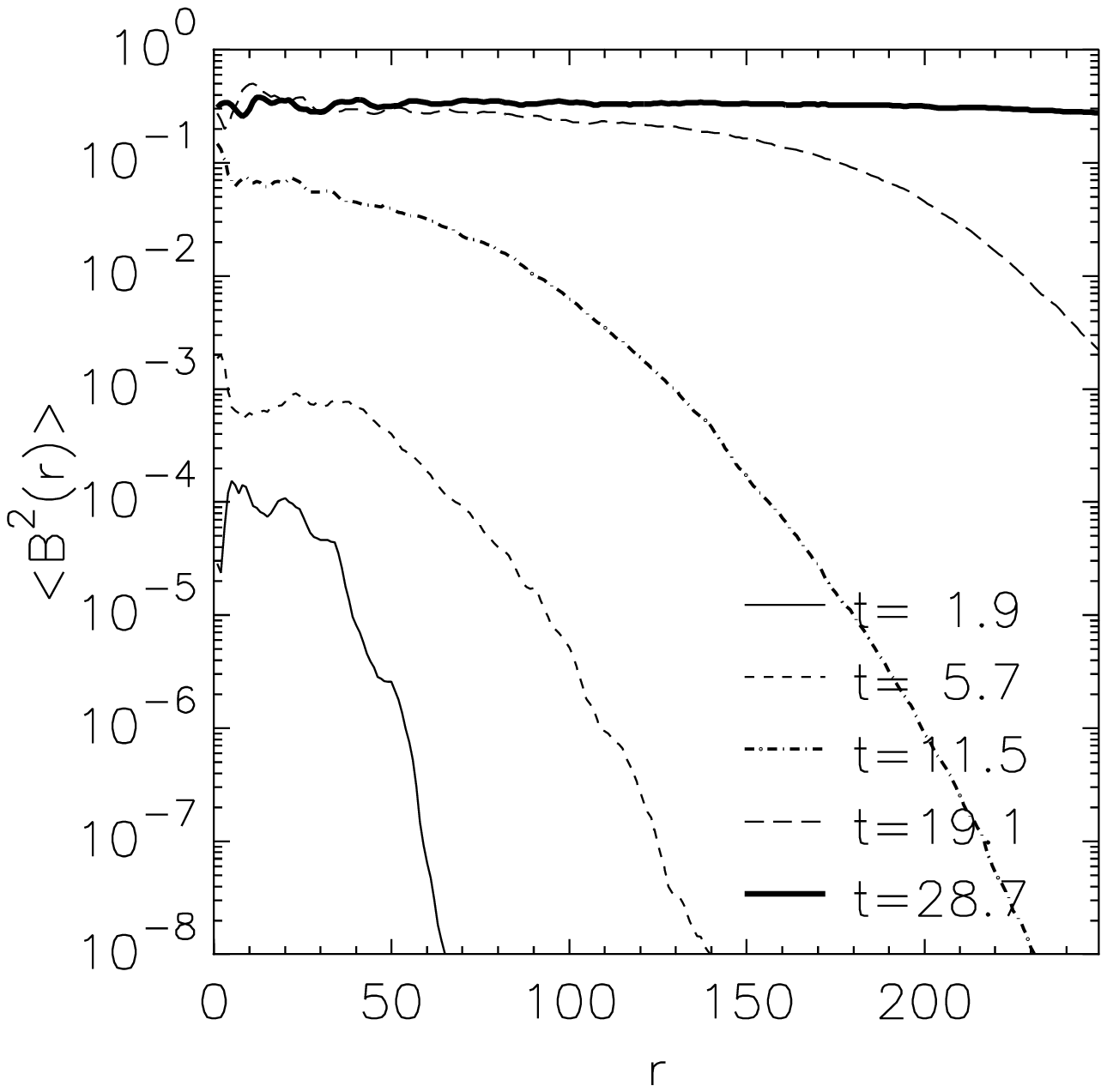}   
\includegraphics[angle=0,width=0.33\textwidth]{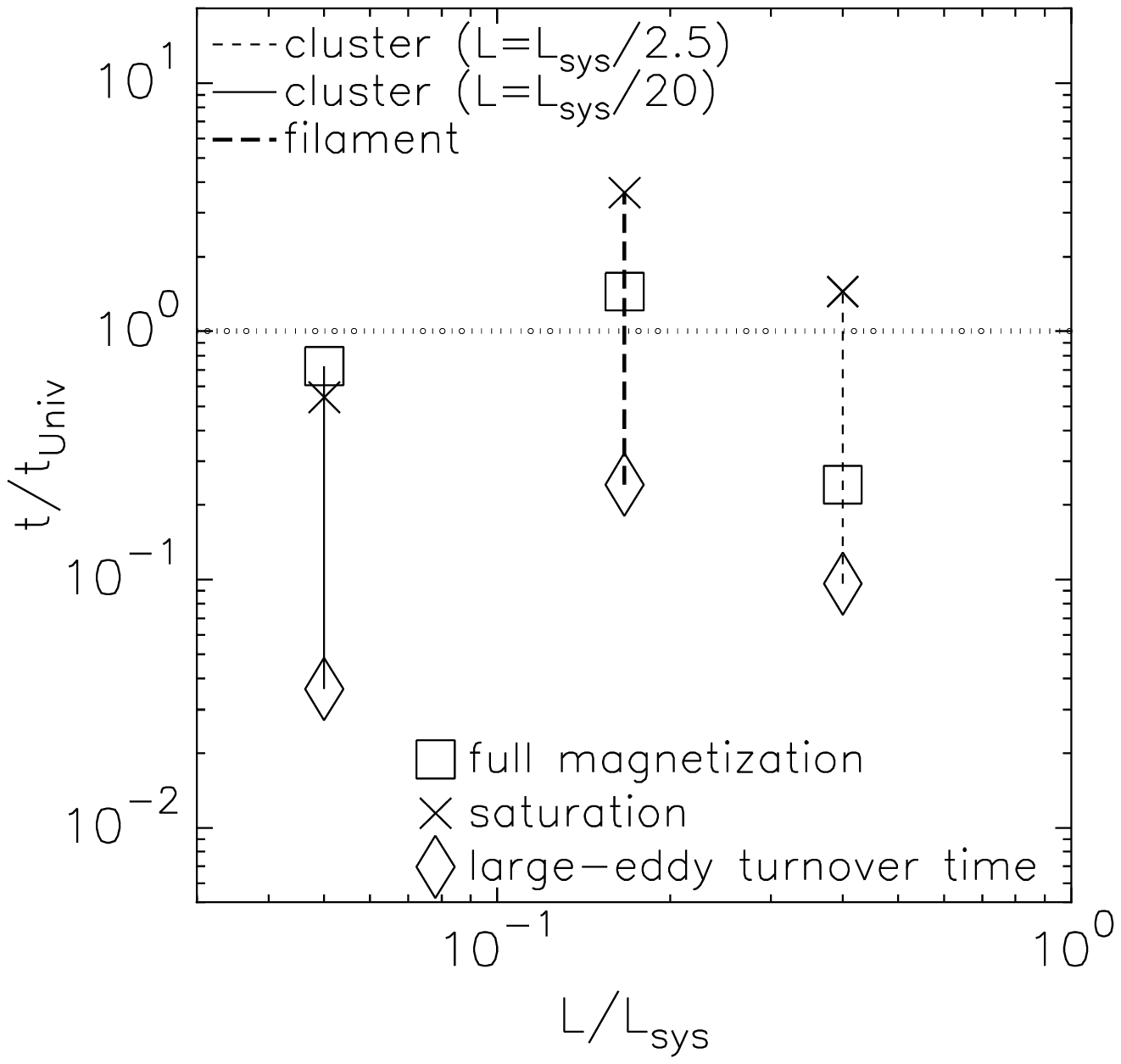}   
\\
\vspace{-5mm}
\caption{(Left) Schematic explanation. 
     In the `magnetized region', 
     magnetic energy density is larger than
     the dissipation-scale kinetic energy density and the magnetic field
     is in either the linear growth stage or the saturation stage. 
     The outer region is 
     not magnetized.
     The lower graph schematically shows 
     the cross-sectional profile of the magnetic energy density.
     (Middle) Profile of $<B^2(r)>$, where $r$ is the distance from the center of the simulation box.
     (Right) Timescales in a filament and clusters.
    The large-eddy turnover time is $\sim L/v$.
     Magnetization timescale ($t_{mag}$) is the timescale for magnetic field to fill
     the whole system and is equal to $\sim (L_{sys}/L)(L/v)$.
     Saturation timescale ($t_{sat}$) is the timescale for the central magnetic field to reach
     the saturation stage and is equal to $\sim$15$L/v$.
 }
\label{fig:exp3}
\end{figure*}

Why is this happening?
Figure \ref{fig:3plots}(b) gives us a useful hint.
Initially $<B_c^2>$, the average  
of $B^2$
inside a sphere of radius 24 (in grid units)
at the center of the simulation box,
shows exponential growth
(see Figure \ref{fig:3plots}(c)).
Then $<B_c^2>$ shows roughly a linear growth for $12 \lesssim t \lesssim 20$ 
(see Figure \ref{fig:3plots}(b)).
After $t \gtrsim 20$, $<B_c^2>$ becomes saturated.
This 3-stage growth is very similar to the growth of a uniform seed magnetic field.
%

\textit{Model} --- 
We note that the beginning of the linear growth stage for $<B_c >$ almost coincides with 
the divergence 
of the behavior of the two fields 
in Figure \ref{fig:3plots}(a).
Based on this observation and 
the 3-stage growth model for a uniform seed magnetic field (see \cite{SchC07,ChoVB09}), 
we can construct a model for a localized seed magnetic field. 

First, when the magnetic field is very weak so that magnetic back-reaction is negligible, 
the magnetic energy grows exponentially,
which is due to stretching of magnetic field lines 
at the dissipation scale.
During this stage, diffusion of magnetic field is very similar to that of the passive scalar.
In our simulation, this stage ends at $t\sim$12.

Second, as the magnetic energy density in the central region reaches  
the kinetic energy density
at the dissipation scale, the exponential growth stage ends due to suppression of stretching 
at the dissipation scale and a slower linear growth stage begins 
in the \textit{central region}.
Note however that the linear growth begins only in the central region.
Magnetic field still grows exponentially in other regions.
As the magnetic field in the central region grows slowly, the distribution of magnetic field may look like
the graph in the left panel of Figure \ref{fig:exp3}. 
The width of the flat central part increases linearly in time. 
This can be understood as
follows. Consider an outer-scale eddy on the boundary of the central region
(see eddy A in the left panel of Figure \ref{fig:exp3}). 
As shown in the figure, some part of the eddy is magnetized and the rest of the eddy
is  
not magnetized. 
We know that magnetic diffusion inside an outer-scale eddy is very fast: it takes roughly
one eddy turnover time for magnetic field to fill the whole eddy \cite{ChoY12}.
Therefore, after approximately one large-eddy turnover time, the whole eddy becomes magnetized.
This way, we have a linear widening (or `expansion') rate
and the speed of expansion is of order $\sim v$, 
the large-scale velocity \footnote{
     An analytic derivation for the expansion rate is possible if we assume
     a Gaussian distribution of magnetic energy density
     (see \cite{MolRS85}; see also \cite{RuzSS89}). 
     In this case, however, the expansion rate
     is faster than ours.
}. 
During this stage the standard deviation of magnetic field distribution, which measures the size of
the flat central region, grows linearly in time and the magnetic energy density will be
proportional to $\sigma^3 <B^2_c> \propto t^4$.

Third, depending on the strength of the seed  field and the ratio $L_{sys}/L$,
   either the magnetic field fills the whole system first or the central magnetic field reaches
   the saturation stage first.
   In our simulation the latter happens first at $t\sim$15-20 (see Figure \ref{fig:3plots}(b)).
   In this case, the standard deviation of magnetic field distribution continues to grow linearly in time
   and the magnetic energy density becomes proportional to $\sim \sigma^3 \propto t^3$.
   Finally, when the magnetic field fills the whole system 
   the growth stage ends (Figure \ref{fig:3plots}(d)).
   On the other hand, if the magnetic field fills the whole system first, 
   the subsequent evolution of magnetic field will be very similar to the linear grow stage
   of a spatially homogeneous/uniform seed field case.

The middle panel of Figure \ref{fig:exp3} shows cross-sectional profile of $B^2$. 
When $t\lesssim 12$, magnetic energy density at the center grows exponentially.
At the same time, the width of the profile increases slowly: $\sigma(t)\propto t^{1/2}$.
At $t\sim$12, the central magnetic energy density becomes comparable to the kinetic energy density at the 
dissipation scale and the exponential growth stops.
After $t\sim$12, the central magnetic field grows slowly and the width of the flat central part
increases.
The overall behavior of the cross-sectional magnetic field profile is consistent with 
our model.

\section{Discussion and Summary}

\textit{Discussion} ---
In this paper, we have found that, during the linear expansion stage, the speed at which
magnetized region \footnote{
  By magnetized region, we mean the region where magnetic energy density is equal to or
  larger than the kinetic energy density at the dissipation scale.
}
 expands is of order $\sim v$.
Therefore, it will take 
\begin{equation}
   t_{mag}\sim L_{sys}/v =(L_{sys}/L)(L/v)
\end{equation}
for magnetic field to fill the whole system.
On the other hand, in the limit of vanishing $\nu~(=\eta)$,
any weak magnetic field at the center can reach the saturation stage
in about 15 large-eddy turnover times, so that we take
\begin{equation}
    t_{sat} \sim 15 (L/v),
\end{equation}
which
is the same as the saturation timescale for a  
uniform seed field case
(see, for example, \cite{Ryu08, ChoVB09, ChoY12}; see also Figure \ref{fig:3plots}(b)).
If $t_{sat} < t_{mag}$ (or, $L_{sys}/L \gtrsim 15$), magnetic field at the center reaches saturation first and
growth of magnetic field ends when it fills the whole system, just as in our simulation.
If $t_{sat}>t_{mag}$ (or, $L_{sys}/L \lesssim 15$),
magnetic field fills the whole system first and continues to grow until it reaches the saturation stage.
In this case, the behavior of the system after $t_{mag}$ will be very similar to 
that of a uniform seed field case.
In general, the growth of magnetic field ends in $t\sim \max( t_{mag}, t_{sat})
=\max( 15, L_{sys}/L ) (L/v)$.

In the large-scale structure of the universe, the driving scale of turbulence
is uncertain (see references in \cite{ChoY12}).
Here, we calculate timescales for three possible examples related
to the large-scale structure of the universe.
We plot the timescales in the right panel of Figure \ref{fig:exp3}. We use
$L_{sys}=3Mpc$, $L=500kpc$, $v=150km/s$ for a filament, 
$L_{sys}=1Mpc$, $L=400kpc$, $v=300km/s$ for a cluster with large-scale driving, and
$L_{sys}=1Mpc$, $L=50kpc$, $v=100km/s$ for a cluster with small-scale driving.
The figure shows that the magnetization timescales for those systems are either  
comparable to or shorter
than the age of the universe.
Therefore, we expect that most volumes in those systems  
are filled with
magnetic fields.
Magnetic field in the filament will be very weak, because it is in early stage
of magnetic field growth.
Magnetic fields in clusters will be relatively strong:
the cluster with large-scale driving ($L=L_{sys}/2.5$)
has almost reached the saturation stage and 
the cluster with small-scale driving ($L=L_{sys}/20$) has already reached the saturation stage.

\textbf{
The right panel of Figure \ref{fig:exp3} implies that it is difficult to tell whether
the origin of the seed magnetic field is cosmological or astrophysical if we observe 
galaxy clusters. 
However, if we observe filaments, it could be possible to
tell the origin of the seed magnetic field because filaments may be partially magnetized 
if the origin is astrophysical and if there are not many sources.
}

\textbf{
In this paper, we have assumed that the viscosity $\nu$ is very small, so that
the Reynolds number ($Lv/\nu$) is very large.
However, there are claims that viscosity in intracluster medium is non-negligible 
\cite{Sch04, Rus04, Rey05,SubSH06}.
If this is the case, what will happen when we inject a weak localized seed magnetic field
at the center of a numerical box?
Of course, it depends on the value of $L/L_{sys}$.
Since the case of  $L\sim L_{sys}$ was discussed in Cho \& Yoo \cite{ChoY12},
let us focus on the case of small $L/L_{sys}$ in this paper.
In particular, let us assume that the seed magnetic field initially grows exponentially and
that the value of $L/L_{sys}$ is so small that
magnetic field at the central region  reaches either a linear growth stage
or a saturation stage before the magnetic field fills the whole system.
If the Reynolds number is less than $\sim O(10^3)$, the outer scale and the dissipation scale
of velocity field
are so close that the growth of the magnetic field  is dominated by the exponential growth stage.
That is, the central magnetic field will grow exponentially most of the time
and, if any, there will be very short linear growth stage \cite{ChoY12}.
During the exponential growth stage, the magnetic field is passively advected by
the outer-scale eddy motions.
Therefore, the behavior of the magnetic field will be very similar to that
of a passive scalar, which means that
diffusion of magnetic field on scales larger than the outer scale is very slow and follows 
a $t^{1/2}$ law.
After the exponential growth stage ends, the magnetic field will show a linear expansion.
The total time for the system to reach a fully magnetized saturation stage will be the sum of
the duration of the exponential growth stage and $\sim L_{sys}/v$.
However, it is not easy to estimate the time because
the duration of the exponential growth stage depends on the strength of the seed magnetic field.
}

\textit{Summary} ---
In this paper, we have shown that magnetic diffusion is very fast in a
turbulent medium even on scales larger than the outer scale of turbulence.
When we inject a weak localized seed magnetic field at the center of 
a turbulent medium,
it initially grows exponentially 
by stretching of magnetic field lines near the dissipation scale %
and fills the outer-scale eddy at the center 
very fast.  %
After filling the outer-scale eddy, diffusion rate slows down ($\sigma \propto t^{1/2}$) because
the magnetic field diffuses over uncorrelated outer-scale eddies.
At this stage, diffusion of the magnetic field is similar to that of a passive scalar.
Then, as the central magnetic field reaches energy equipartition with dissipation-scale 
velocity field, the standard deviation of the magnetic field begins to show a faster growth rate.
At this stage, the speed at which the magnetized region `expands' is of order $\sim v$,
which enables a full magnetization of the system in a timescale of $\sim L_{sys}/v$.

In the limit of $\nu$ (=$\eta$) $\rightarrow 0$,
growth of a localized seed magnetic field ends in $\sim \max( 15, L_{sys}/L )(L/v)$.
Note that earlier studies \cite{Ryu08,ChoVB09} 
have shown that turbulence can efficiently amplify a spatially uniform or homogeneous
seed field, the timescale of which is $\sim 15(L/v)$.
Therefore, unless $L_{sys}/L$ is extremely large,
we can conclude that turbulence can amplify any shape of seed magnetic field
very fast.

\acknowledgements
This research was supported by National R \& D Program through 
the National Research Foundation of Korea (NRF) 
funded by the Ministry of Education, Science and Technology (No. 2012-0002798).


\begin{thebibliography}{99}   
\bibitem[Name (2013)]{Kro94} P.~P.~Kronberg, Rep. Prog. Phys., 57, 325 (1994)

\bibitem[Name (2013)]{KulZ08} R.~M.~Kulsrud, \& E.~G.~Zweibel,  
           Reports on Progress in Physics, 71, 046901 (2008)

\bibitem[Name (2013)][Rees(1987)]{Rees87}  M.~J.~Rees, 
         Royal Astronomical Society, Quarterly Journal, 28, 197 (1987)

\bibitem[Name (2013)][Brandenburg(1996)]{BraEO96} A.~Brandenburg, K.~Enqvist, \& P.~Olesen,
                Phys. Rev. D, 54, 1291 (1996)
\bibitem[Name (2013)]{Bat50} G.~Batchelor, Proc. R. Soc. London A, 201, 405 (1950)

\bibitem[Name (2013)]{Kaz68} A.~P.~Kazantsev, Soviet Phys.-JETP Lett., 26, 1031 (1968)

\bibitem[Name (2013)]{KulA92} R.~Kulsrud,  \& S.~Anderson,, \apj, 396, 606 (1992)

\bibitem[Name (2013)]{Kul97}  R.~M.~Kulsrud, R.~Cen, J.~P.~Ostriker, \&D.~ Ryu,   \apj, 480, 481 (1997)

\bibitem[Name (2013)]{ChoV00} J.~Cho, \& E.~T.~Vishniac,  \apj, 538, 217 ( 2000)

\bibitem[Name (2013)]{Sch04} A.~A.~Schekochihin, S.~C.~Cowley, S.~F.~Taylor, J.~L.~Maron, \& J.~C.~McWilliams,  
                 \apj, 612, 276 (2004)

\bibitem[Name (2013)]{BraS05} A.~Brandenburg,  \& K.~Subramanian,   Phys. Reports, 417, 1 (2005)

\bibitem[Name (2013)]{SubSH06}  K.~Subramanian, A.~Shukurov,  \& N.~Haugen,   \MNRAS, 366, 1437 (2006)

\bibitem[Name (2013)]{SchC07} A.~Schekochihin,  \& S.~Cowley, in 
       {\it Magnetohydrodynamics - Historical evolution and trends},
        eds. by S. Molokov, R. Moreau, \& H. Moffatt
       (Berlin; Springer), p.~85. (2007) (astro-ph/0507686)


\bibitem[Name (2013)]{Ryu08} D.~Ryu, H.~Kang, J.~Cho, \& S.~Das,  Science, 320, 909 (2008)

\bibitem[Name (2013)]{ChoVB09} J.~Cho, E.~T.~Vishniac,  
                    A.~Beresnyak,  A.~Lazarian, \&D.~ Ryu,  \apj, 693, 1449 (2009)

\bibitem[Name (2013)]{ChoY12} J.~Cho,  \& H.~Yoo,   \apj, 759, 91 (2012)

\bibitem[Name (2013)]{Rus04} M.~Ruszkowski, M.~Br\"uggen, \& M.~Begelman, \apj, 611, 158 (2004)
\bibitem[Name (2013)]{Rey05} C.~Reynolds, B.~McKernan, A.~Fabian, J.~Stone, \& J.~Vernaleo, \MNRAS, 357, 242 (2005)


\bibitem[Name (2013)]{MolRS85} S.~Molchanov,  A.~Ruzmaikin,  \& D.~Sokolov,   Sov. Phys. Usp., 28, 307 (1985)

\bibitem[Name (2013)]{RuzSS89}  A.~Ruzmaikin, D.~Sokoloff, \& A.~Shukurov,   \MNRAS, 241, 1 (1989)


\end{thebibliography}
\end{document}